\documentclass[journal,comsoc]{IEEEtran}
\usepackage{cite}
\usepackage{graphicx}
\usepackage{amsmath}
\usepackage{times}
\usepackage{latexsym}
\usepackage{graphicx}
\usepackage{bm}
\usepackage{amssymb}
\usepackage[center]{caption2}
\usepackage{stfloats}
\usepackage{cases}
\usepackage{subfigure}
\usepackage{array}
\usepackage{setspace}
\usepackage{fancyhdr}
\usepackage{citesort}
\usepackage{colortbl}
\usepackage{CJK}
\usepackage{tikz}
\usepackage{float}
\usepackage{color}
\usepackage{bbding}

\usepackage{bm}
\usepackage{mathrsfs}
\usepackage{diagbox}

\renewcommand{\QED}{\QEDopen}

\allowdisplaybreaks[4]

\renewcommand{\baselinestretch}{0.97}

\newcaptionstyle{mystyle2}{%
\captionlabel $.$ \, \captiontext \par} \captionstyle{mystyle2}
\renewcommand\thesubsection{\Alph{subsection}}

\begin{document}
\title{Online Deep Neural Network for Optimization in Wireless Communications}
\author{\authorblockN{Jiabao Gao, Caijun Zhong, Geoffrey Ye Li, and Zhaoyang Zhang
\thanks{J. Gao, C. Zhong, and Z. Zhang are with the College of Information Science and Electronic Engineering, Zhejiang University, Hangzhou, China (Email:  \{gao\_jiabao, caijunzhong, ning\_ming\}@zju.edu.cn). Geoffrey Ye Li is with the Faculty of Engineering, Department of Electrical and Electronic Engineering, Imperial College London, England (Email: Geoffrey.Li@imperial.ac.uk).}
}}
\maketitle

\begin{abstract}
Recently, deep neural network (DNN) has been widely adopted in the design of intelligent communication systems thanks to its strong learning ability and low testing complexity. However, most current offline DNN-based methods still suffer from unsatisfactory performance, limited generalization ability, and poor interpretability. In this article, we propose an online DNN-based approach to solve general optimization problems in wireless communications, where a dedicated DNN is trained for each data sample. By treating the optimization variables and the objective function as network parameters and loss function, respectively, the optimization problem can be solved equivalently through network training. Thanks to the online optimization nature and meaningful network parameters, the proposed approach owns strong generalization ability and interpretability, while its superior performance is demonstrated through a practical example of joint beamforming in intelligent reflecting surface (IRS)-aided multi-user multiple-input multiple-output (MIMO) systems. Simulation results show that the proposed online DNN outperforms conventional offline DNN and state-of-the-art iterative optimization algorithm, but with low complexity.
\end{abstract}


\section{INTRODUCTION}
Recently, thanks to its strong learning ability and low testing complexity, deep neural network (DNN) has made great success in optimization problems in wireless communication, such as channel estimation\cite{CE1}, beamforming\cite{MultiUser}, signal detection\cite{SD1}, resource allocation\cite{B_B}, etc. The existing DNN-based optimization methods can be mainly divided into two categories, to improve algorithm performance and efficiency, respectively. For the first category, black-box DNNs are used to directly learn the input-to-output mapping\cite{CE1,MultiUser}, or techniques, such as deep unfolding, are used to exploit the advantages of both deep learning and conventional iterative optimization algorithms\cite{SD1}. Besides, the optimization objectives are expressed in the form of the monotonically decreasing energy functions of Hopfield neural networks in \cite{Hopfield} so that the objective is optimized as the network evolves. As for the second category, the pruning policy is learned in \cite{B_B} to accelerate the branch and bound algorithm while the complex objective function is approximated by a network in \cite{ApproximateObjective} to solve the problem with simple optimization techniques.

In most current DNN-based wireless optimization works, the network is first trained offline with a large number of samples to minimize the average loss over the entire dataset, and the network parameters are fixed during online testing. In spite of high theoretical performance according to the universal approximation theory, the actual performance of offline DNN can be limited by inadequate training and local minima. Consequently, in many complex wireless communication problems, conventional algorithms are still superior in performance, and the advantages of DNN mainly lie in lower complexity\cite{MultiUser}. Besides, the performance degradation of the DNN trained offline is very common when input distribution changes during online testing\cite{MultiUser}, and the limited generalization ability hinders the application of DNN in fast changing environments. Last but not least, DNN is often regarded as a black box with unexplainable parameters in data-driven methods, therefore not suitable for tasks with strict reliability requirements.

To address the above issues in the current offline DNN-based methods, we propose a novel online DNN-based approach in this article to solve general optimization problems in wireless communication, where a dedicated DNN is trained for each data sample. Specifically, the optimization variables and the objective function are treated as network parameters and loss function, respectively. Then, the decrease of loss through network training is equivalent to the solving process of the optimization problem. The strong generalization ability and interpretability of the proposed approach can be easily understood based on its online optimization nature and meaningful network parameters. Furthermore, a practical example is provided to facilitate a better understanding of the proposed approach and illustrate its superiority. In the joint beamforming problem in intelligent reflecting surface (IRS)-aided multi-user multiple-input multiple-output (MIMO) systems, we demonstrate that the proposed online DNN achieves better performance than conventional offline DNN and state-of-the-art iterative optimization algorithm, but with low complexity.


\section{Online DNN for general optimization}
In this section, the general proposed framework is elaborated from four aspects, namely network modeling, constraint elimination, parameter initialization, and network training.

\subsection{Network Modeling}
Consider the following unconstrained optimization problem:
\begin{equation}
\min\limits_{\bm{x}} f(\bm{a},\bm{x}),
\end{equation}
where $\bm{a}$ denotes known parameters, $\bm{x}$ denotes optimization variables, and $f$ denotes the objective function. 

\begin{figure}[htbp]
\centering
\includegraphics[width=0.50\textwidth]{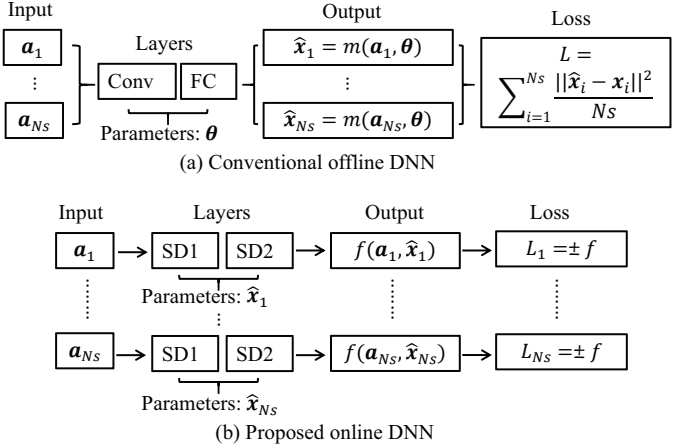}
\caption{The frameworks of the conventional offline DNN and the proposed online DNN to solve optimization problem (1).}
\label{framework}
\end{figure}

Both the conventional offline DNN-based method and the proposed online DNN-based approach can be used to solve the above optimization problem. Fig. \ref{framework} illustrates the frameworks of two methods and their main components are compared as follows to highlight the novelty of the proposed approach:
\begin{itemize}
\item {\bf Input}: For both methods, known parameters $\bm{a}$ that contain available information are treated as network input.
\item {\bf Layers \& Parameters}: The conventional offline DNN typically consists of convolutional (Conv) and fully-connected (FC) layers with unexplainable parameters $\bm{\theta}$ while the proposed online DNN adopts self-defined (SD) layers where the estimations of optimization variables $\hat{\bm{x}}$ are treated as parameters and the forward computation is customized according to the signal flow to obtain $f$.
\item {\bf Output}: The output of the conventional offline DNN is the estimations of optimization variables $\hat{\bm{x}}=m(\bm{a},\bm{\theta})$, where $m$ denotes the unexplainable mapping function parameterized by $\bm{\theta}$. For the proposed online DNN, the output is the optimization objective $f(\bm{a},\hat{\bm{x}})$. 
\item {\bf Loss function}: In the conventional offline DNN, consider supervised learning, the mean-squared error between the network prediction and the label is commonly used as the loss function, $L$. In the proposed online DNN, $f$ is a reasonable choice for $L$ since its reduction through training is equivalent to the solving process of the optimization problem. Since no label is required, methods using this kind of loss function are usually called unsupervised learning-based approaches in recent literature\cite{MultiUser,GraphNN}. Notice that, in maximization problems, $L$ should be $1/f$ or $-f$ so that the reduction of loss is equivalent to the maximization of the objective function.
\end{itemize}

As demonstrated by Fig. \ref{framework}, the conventional approach trains a common network offline with multiple data samples while the proposed approach trains a dedicated network online for each new sample. Therefore, there is no so-called testing stage in the online DNN since $\hat{\bm{x}}$ are obtained at network parameters rather than output, and the generalization problem does not exist at all. Besides, the meaningful parameters make the online DNN highly interpretable.

\subsection{Constraint Elimination}
DNN parameters are usually unconstrained and can take arbitrary values in the entire real space. However, for those optimization problems in wireless communications, the optimization variables, $\bm{x}$, are subject to various constraints, which results in a feasible region $\mathcal{X}$.

To implement the proposed DNN with constrained variables, an intuitive method is to eliminate the constraints and transform the constrained optimization problems to unconstrained ones. There are several standard methods on constraint elimination. Some integrate the constraints into the objective function, such as the Lagrangian multiplier method and the penalty function method while others maintain the feasibility of the solution through projection operation, such as the projected gradient descent algorithm\cite{constraint_elimination}. Nevertheless, these methods suffer from complex mathematical derivation, ill-conditioned problems, and slow convergence.

In this article, we use the technique of reparameterization. Specifically, for $\bm{x}\in \mathcal{X}$, if we can find a differentiable transform function $g$ to express $\bm{x}$ in the form of a set of unconstrained variables $\bm{x}'$, i.e., $\bm{x}=g(\bm{x}')$ and the feasible region of $\bm{x}'$ is the entire real space, then we can treat $\bm{x}'$ as network parameters instead of $\bm{x}$. During training, the gradients of the loss function with respect to $\bm{x}'$ can be obtained by the chain rule, i.e., $\frac{dL}{d\bm{x}'}=\frac{dL}{d\bm{x}}\frac{d\bm{x}}{d\bm{x}'}$. After training, $\hat{\bm{x}}$ can be readily recovered by $g(\hat{\bm{x}'})$. Next, we provide the transforms and unconstrained counterparts of optimization variables of most common constraints in wireless communications:
\begin{itemize}
\item {\bf Complex constraint}: In most communication problems, the optimization variables are complex numbers. If an optimization variable $x\in\mathbb{C}$, then the unconstrained counterparts are its real and imaginary parts $x'_r$ and $x'_i$, and the transform is $x=x'_r+jx'_i$.
\item {\bf Unit modulus constraint}: When IRS or phase shifters are used, phases of components have unit modulus. If $x\in\mathbb{C}$ and $|x|=1$, then the unconstrained counterpart is its argument $\phi$, and the transform is $x=e^{j\phi}$.
\item {\bf Box constraint}: If $a\le x\le b$, then the transform is $x=a+(b-a)\text{Sigmoid}(x')$, where the value of $\text{Sigmoid}(x')=1/(1+e^{-x'})$ is between 0 and 1.
\item {\bf Maximum power constraint:}
If $\bm{x}\in \mathbb{R}^K$ satisfies $\sum_{k=1}^Kx_k\le P$, then the unconstrained counterparts are the power unconstrained version $\bm{x}'$ and a power scaler $c$. The transform is $\bm{x}=\bm{x}'/\sum_{k=1}^Kx'_{k}\times P\times \text{Sigmoid}(c)$. When beamforming is considered with multi-antenna transmitters, the transform is similar, as will be introduced later in the given example.
\item {\bf Linear equality constraint:} If $\bm{x}\in \mathbb{R}^K$ satisfies $\bm{Ax}=\bm{b}$, where $\bm{A}\in\mathbb{R}^{M \times K}$ is full row rank and $M<K$, i.e., there are infinite feasible solutions of $\bm{x}$. Then, the transform is $\bm{x}=\bm{Fx}'+\bm{x}_0$, where $\bm{x}_0$ is a special solution that satisfies $\bm{Ax}_0=\bm{b}$, e.g., $\bm{x}_0=\bm{A}^\dagger\bm{b}$ with $\dagger$ denoting pseudo inverse, and $\bm{F}\in\mathbb{R}^{K\times(K-M)}$ is the zero space of $\bm{A}$, which satisfies $\bm{AF}=\bm{0}$ and can be obtained by the $null$ function in Matlab or Python.
\item {\bf Linear inequality constraint:} If $\bm{x}\in \mathbb{R}^K$ satisfies $\bm{Ax}\le \bm{b}$, then the transform is $\bm{x}=\bm{Fx}'+\bm{A}^\dagger(\bm{b}-\bm{\mu})$, where $\bm{\mu}=e^{\bm{\mu}'}>0$ denotes the introduced set of slack variables.
\end{itemize}

\subsection{Parameter Initialization}
Before training, network parameters need to be properly initialized first, which is especially important in non-convex optimization problems. One simple method is to use random generalization. Besides, by initializing and training multiple times and selecting the best one, the performance can be improved and stabilized, albeit at the cost of higher complexity.

In fact, high quality initializations can also be found without much complexity overhead by exploiting expert knowledge. For instance, we can initialize with sub-optimal solutions obtained by low-complexity baseline algorithms. Or, in low-mobility scenarios, the channels are highly time-correlated, so the current initialization can inherit from previously optimized parameters or even be predicted by autoregressive models.

\subsection{Network Training}
After parameter initialization, the training process begins. During training, network parameters can be optimized by popular DNN optimizers. Specifically, in each training iteration, the network first executes forward computation to obtain the loss, and then executes backward computation to compute the gradients of the loss function with respect to all network parameters, which is efficiently implemented by mainstream deep learning libraries. Based on the gradients and the learning rate, network parameters are updated correspondingly. Multiple iterations are required to train the network to convergence. The learning rate, which is the only hyper-parameter in the proposed online DNN, has to be carefully configured to improve training efficiency. After the training process is completed, final results of optimization variables can be readily recovered based on the network parameters and the corresponding transforms.

\subsection{Relationship with Classic Gradient Descent}
Actually, the proposed approach is equivalent to the classic gradient descent algorithm theoretically. However, conventional manual derivation of gradients or symbolic differentiation suffers from swelling expressions and low computation efficiency, while the proposed novel neural network-based implementation benefits from automatic differentiation and paves the way for fast and universal applications of gradient descent in practical optimization problems. Despite the simplicity of the core algorithm, surprisingly good results can be achieved sometimes, such as the example given in the next section.

\section{Online DNN for Joint Beamforming in IRS-aided Multi-user MIMO systems}
To facilitate a better understanding of the proposed approach and illustrate its superiority, we elaborate joint beamforming in IRS-aided multi-user MIMO systems as an example.

\subsection{System Model and Problem Formulation}
\begin{figure}[htbp]
\centering
\includegraphics[width=0.50\textwidth]{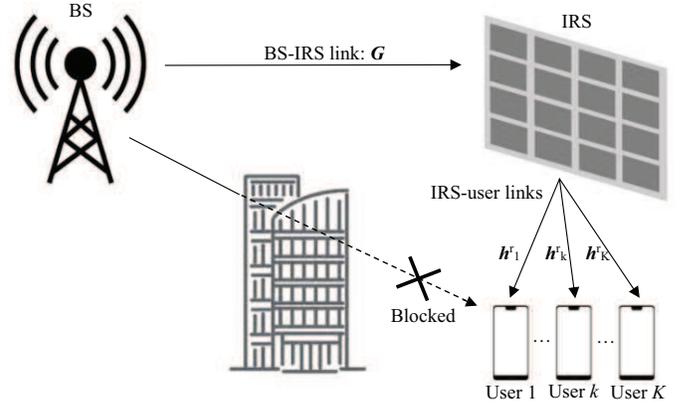}
\caption{IRS-aided multi-user MIMO system}
\label{system}
\end{figure}

Consider the IRS-aided multi-user MIMO system illustrated in Fig. \ref{system}, where the BS with $M$ antennas serves $K$ single-antenna users with the aid of an IRS with $N$ reflecting elements. The direct links between the BS and users are assumed to be blocked. The received signal at the $k$-th user can be written as
\begin{align}
\label{transmission_model}
y_k=\bm{h}^r_k\bm{\Theta Gx}+n_k,
\end{align}
for $k=1, \cdots, K,$ where $\bm{h}^r_k\in {\mathbb C}^{1\times N}$ and $\bm{G}\in {\mathbb C}^{N\times M}$ denote the channels of the $k$-th IRS-user link and the BS-IRS link, respectively. The phase shift matrix of IRS is defined as $\bm{\Theta}\triangleq \text{diag}([\theta_1,...,\theta_N])$, where $|\theta_n|=1$ is the phase shift of the $n$-th reflecting element, $\text{diag}(\cdot)$ denotes the diagonalization operation, and $\bm{x}=\sum_{k=1}^K\bm{w}_{k}s_k$ is the transmit signal at the BS, where $\bm{w}_{k}\in{\mathbb C}^{M\times 1}$ and $s_k$ satisfying $\mathbb{E}\{s_ks_k^*\}=1$ denote the transmit beamforming vector and the information symbol for the $k$-th user, respectively. Besides, $n_k\sim \mathcal{CN}(0,\sigma^2)$ denotes the noise at the $k$-th user with variance $\sigma^2$.

Define $\bm{W} \triangleq [\bm{w}_1, ..., \bm{w}_k]$ and $\bm{H}^r\triangleq [\bm{h}^{rT}_1, ..., \bm{ h}^{rT}_K]^T$, the effective channel matrix is defined as $\bm{H}\triangleq \bm{H}^r\bm{\Theta G}\in {\mathbb C}^{K\times M}$. Then, the received signal-to-interference-plus-noise ratio (SINR) at the $k$-th user can be expressed as
\begin{align}
    {\gamma_k}=\frac{\bm{w}_k^H\bm{H}_{k*}^H\bm{H}_{k*}\bm{w}_k}{J_k},
\end{align}
for $k=1, \cdots, K,$ where $J_k\triangleq\sigma ^2+\sum_{i=1,i\ne k}^K\bm{w}_i^H\bm{H}_{k*}^H\bm{H}_{k*}\bm{w}_i$ is the energy of interference plus noise at the $k$-th user and $\bm{H}_{k*}$ denotes the $k$-th row vector of $\bm{H}$. We aim to maximize the sum rate of all users $\mathcal R$, by jointly optimizing the transmit beamforming matrix $\bm{W}$ and the IRS phase shift matrix $\bm{\Theta}$. The optimization problem is given by
\begin{subequations}
\begin{align}
\label{opt_problem}
\max\limits_{\bm{\Theta},\bm{W}}\quad& {\mathcal R}=\sum_{k=1}^K\log(1+{\gamma_k})\\
\text{s.t.}\quad&\sum_{k=1}^K\bm{w}^H_k\bm{w}_k\leq P_{max},\\
&|\theta_i|=1, \forall i=1, 2, ..., N,
\end{align}
\end{subequations}
where (4b) is the transmit power constraint and $P_{max}$ denotes the maximum transmit power at the BS, while (4c) is the unit modulus constraint of phase shifts of IRS reflecting elements.

\subsection{Detailed Designs of the Proposed Online DNN}
According to (2), (3), and (4a), we can easily find the counterparts of the main components of the general framework in this specific problem. Apparently, $\bm{G}$, $\bm{H}^r$, and $\sigma^2$ make up known parameters $\bm{a}=\{\bm{G}$, $\bm{H}^r, \sigma^2\}$, while $\bm{\Theta}$ and $\bm{W}$ make up optimization variables $\bm{x}=\{\bm{\Theta}, \bm{W}\}$ and $R$ is the objective function $f$. Besides, the unconstrained counterparts of $\bm{\Theta}$ and $\bm{W}$ as well as proper transforms are required to handle constraints (4b) and (4c).
\begin{figure*}[htbp]
\centering
\includegraphics[width=0.7\textwidth]{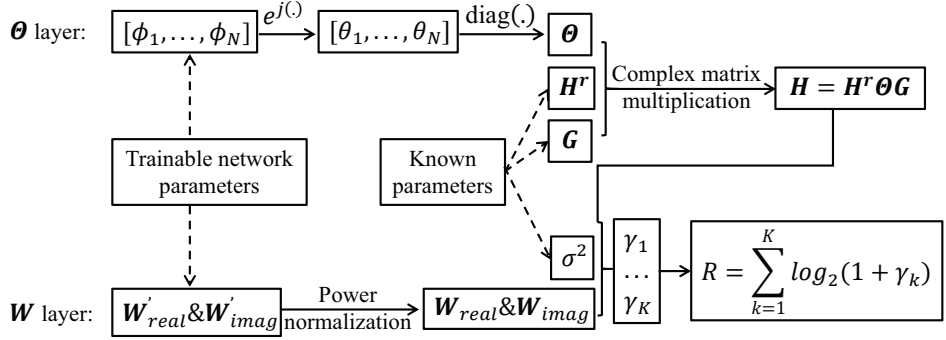}
\caption{Network architecture for joint beamforming in IRS-aided multi-user MIMO systems.}
\label{network}
\end{figure*}

The detailed network structure is illustrated in Fig. \ref{network}. It is straightforward to implement two layers representing $\bm{\Theta}$ and $\bm{W}$, respectively. First of all, $\bm{G}$ and $\bm{H}^r$ are input into the $\bm{\Theta}$ layer. Inside the $\bm{\Theta}$ layer, the arguments of phase shifts of IRS reflecting elements, $\phi_1,\cdots,\phi_N$, are defined as $N$ trainable network parameters. The forward computation first transforms $\phi$ to $\theta$ by $\theta_i=e^{j\phi_i},i=1,\cdots,N$. Then, the effective channel matrix $\bm{H}$ is computed based on $\bm{G}$, $\bm{H}^r$, and $\bm{\Theta}$. Afterwards, $\bm{H}$ output by the $\bm{\Theta}$ layer flows into the $\bm{W}$ layer together with $\sigma^2$. Inside the $\bm{W}$ layer, the power unconstrained real and imaginary parts of the transmit beamforming matrix $\bm{W}$'s elements $\bm{W}'_{real}$ and $\bm{W}'_{imag}$, are defined as $2KM$ trainable network parameters. The forward computation first realizes the transform of power normalization by $\bm{W}_{real\&imag}=\bm{W}'_{real\&imag}/\sqrt{\sum_{k=1}^K\bm{w}'^H_k\bm{w}'_k}\times\sqrt{P_{max}}$. Then, the SINRs of users $\gamma_k,k=1,\cdots,K$, are computed based on $\bm{W}$, $\bm{H}$, and $\sigma^2$. Eventually, the sum rate of all users, $R$, can be readily computed and the loss function defined as $L=-R$ is used for network training.

\subsection{Simulation Results}
Next, the superiority of the proposed approach is validated through simulation. Adopt the Rician channel model, the channels of the BS-IRS link and the $k$-th IRS-user link are 
\begin{align}
    \bm{G}= L_1(\sqrt{\frac{\epsilon}{\epsilon+1}}\bm{a}_N(\nu)\bm{a}_M(\phi)^H+\sqrt{\frac{1}{\epsilon+1}}\overline{\bm{G}}),\\
    \bm{h}_k^r= L_{2,k}(\sqrt{\frac{\epsilon}{\epsilon+1}}\bm{a}_N(\zeta_k)+\sqrt{\frac{1}{\epsilon+1}}\overline{\bm{h}_k^r}),
\end{align}
where $L_1$ and $L_{2,k}$ are path-losses in dB calculated as ${35.6+22.0\text{lg}(d)}$ with $d$ denoting the distance, $\bm{a}_M$ and $\bm{a}_N$ are the steering vectors of uniform linear array at the BS and the IRS, respectively, while $\nu$, $\phi$ and $\zeta_k$ are angular parameters. The Rician factor $\epsilon$ is set to 10, while $\overline{\bm{G}}$ and $\overline{\bm{h}_k^r}$ are non-line-of-sight components following $\mathcal{CN}(0,1)$. The distance between the BS and the IRS is fixed to $200$ m, users are uniformly distributed in a circle $30$ m away from the IRS with a radius of $10$ m, and $P_{max}/\sigma^2$ is fixed to 20 dB.

\begin{figure}[!htb]
\centering
\includegraphics[width=0.45\textwidth]{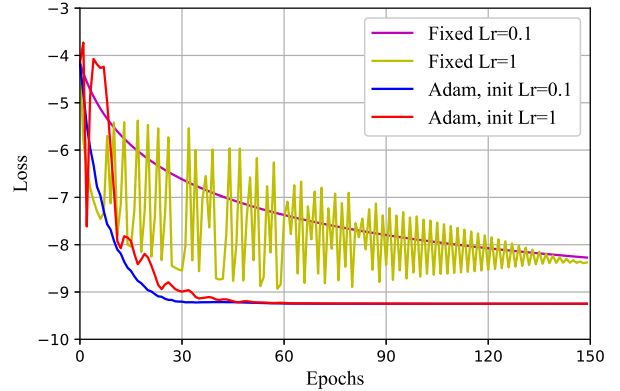}
\caption{Convergence process of an exemplary sample.}
\label{impact_of_training}
\end{figure}

Firstly, the impact of learning rate configuration is investigated. In the considered problem, the proposed DNN already works well with randomly initialized $\bm{\Theta}$ and $\bm{W}$. The training process terminates when the loss does not decrease in 25 consecutive iterations. Fig. \ref{impact_of_training} illustrates the convergence process of an exemplary sample when $M=8, K=4$ and $N=64$. As we can see, when the learning rate is fixed, a large learning rate can cause severe oscillation while a small learning rate can lead to slow convergence. In contrast, Adam is more superior in terms of convergence speed and performance and is less sensitive to the initial learning rate thanks to its adaptive adjustment of learning rate. Therefore, we adopt Adam with initial learning rate 0.1. Notice that, the usage of the advanced Adam optimizer originally developed in the area of deep learning benefits from our clever modeling of the optimization problem as a DNN.

\begin{figure}[!htb]
\centering
\includegraphics[width=0.45\textwidth]{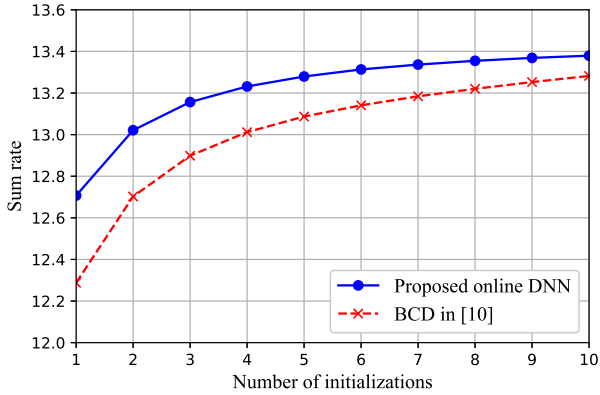}
\caption{Impact of the number of initializations.}
\label{impact_of_init}
\end{figure}

The impact of initialization is illustrated in Fig. \ref{impact_of_init}, where $M=8, K=4$ and $N=128$. The best result obtained by running with multiple initializations are kept. The state-of-the-art block coordinate descent (BCD) algorithm\cite{baseline} is selected as a baseline. For BCD, random phase shifts and weighted minimum mean-squared error (WMMSE) beamforming based on the effective channels serve as the initializations of $\bm{\Theta}$ and $\bm{W}$, respectively, and the algorithm stops when the change of sum rate between two consecutive iterations is less than 1e-5. As we can see, the performance of both the proposed approach and BCD improves with the number of initializations at the cost of increased complexity, while the proposed approach consistently outperforms BCD, which can be attributed to the simultaneous update of all parameters. Besides, the performance gap decreases with the number of initializations due to BCD's larger performance variance of different initializations.

\begin{figure}[!htb]
\centering
\includegraphics[width=0.45\textwidth]{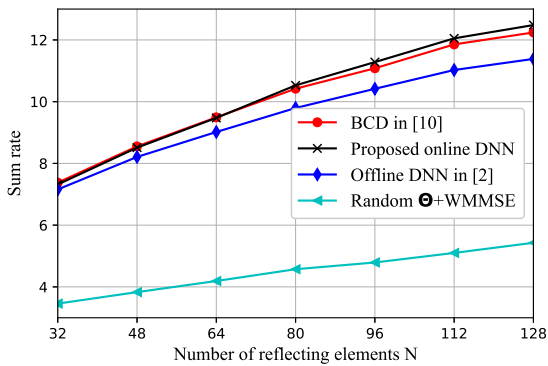}
\caption{Impact of the number of reflecting elements $N$.}
\label{impact_of_N}
\end{figure}

To save running time, we consider single initialization next.
The impact of the number of reflecting elements $N$ is illustrated in Fig. \ref{impact_of_N}, where $M=8$ and $K=4$. The offline DNN-based approach proposed in \cite{MultiUser} with unsupervised training is also compared to highlight the superiority of the proposed online DNN. As we can see, the proposed approach achieves similar performance as BCD when $N$ is small, while when $N\ge80$, the proposed approach outperforms BCD and the performance gap increases with $N$. It is because the probability of BCD converging to a worse local optimum than the proposed approach is higher in systems with larger scales. Nevertheless, both the proposed approach and BCD outperform the offline DNN  with various $N$. Notice that, for the offline DNN, performance degradation can happen when channel parameters changes\cite{MultiUser}, which does not exist in the proposed approach thanks to its online optimization nature.

\subsection{Complexity Analysis}
The complexity of the BCD algorithm is $\mathcal{O}(I_O(2KNM+KM^2+K^2N^2))$, where $I_O$ denotes the number of iterations\cite{baseline}. The complexity of the offline DNN proposed in \cite{MultiUser} is $\mathcal{O}(KNM)$. As for the proposed approach, the complexity is $\mathcal{O}(I_E(C_{F}+C_{B}))$, where the forward computation complexity $C_F$ is $\mathcal{O}(KNM+K^2M+KM)$, the predominant backward computation complexity $C_B$ is $\mathcal{O}(K^2NM+K^3M^2)$, and $I_E$ denotes the number of training iterations. Since usually $N\gg M$ and $N\gg K$, the proposed approach has lower per-iteration complexity than BCD. Besides, the proposed approach also requires less iterations to converge in experiments. Although the offline DNN has the lowest complexity, its performance is apparently inferior and the generalization and interpretability issues also hinder its practical applications. From a certain point of view, the proposed online DNN achieves the best performance-complexity tradeoff in the considered problem. To make the comparison more intuitive, running time on the same CPU is further shown in Table \ref{complexity}. The proposed approach runs much faster than BCD, especially in large scale systems.

Notice that, another benefit of the proposed approach is the acceleration thanks to its DNN-based structure, including the efficient implementation of matrix calculation and the gradient backpropagation algorithm in deep learning libraries, as well as the usage of dedicated hardware like GPU for parallel acceleration. Besides, in some special problems, the decomposition of loss calculation into independent blocks for further acceleration is a future direction worth investigating.
\begin{table}[!htb]
\begin{tabular}{|c|c|c|c|c|}
\hline
\diagbox{$M,N,K$}{Method} & Proposed & BCD & Offline DNN\\ \hline
4,64,2 & 0.196 & 0.383 & 0.002\\ \hline
8,64,2 & 0.220 & 0.816 & 0.003\\ \hline
8,128,2 & 0.381 & 6.034 & 0.009\\ \hline
8,128,4 & 0.587 & 9.871 & 0.011\\ \hline
\end{tabular}
\centering
\caption{Average running time in seconds.}
\label{complexity}
\end{table}

\section{CONCLUSION}
\label{conclusion}
In this article, we have developed a novel online DNN-based approach to solve general optimization problems in wireless communications. By treating the optimization variables and the objective function as network parameters and loss function, respectively, the optimization problem can be equivalently solved through network training. The proposed approach has strong generalization ability and interpretability, and outperforms conventional offline DNN and iterative optimization algorithm with low complexity in a practical example.


\begin{thebibliography}{99}

\bibitem{CE1}
J. Gao, M. Hu, C. Zhong, G. Y. Li and Z. Zhang, ``An attention-aided deep learning framework for massive MIMO channel estimation," \emph{IEEE Trans. Wireless Commun.}, Early Access.
\bibitem{MultiUser}
H. Song, M. Zhang, J. Gao, and C. Zhong, ``Unsupervised learning based joint active and passive beamforming design for recongurable intelligent surfaces aided wireless networks," \emph{IEEE Commun. Lett.}, vol. 25, no. 3, pp. 892--896, Mar. 2021.
\bibitem{SD1}
H. He \emph{et al}., ``Model-driven deep learning for MIMO detection," \emph{IEEE Trans. Signal Process.}, vol. 68, pp. 1702--1715, Feb. 2020.
\bibitem{B_B}
M. Lee, G. Yu, and G. Y. Li, ``Learning to branch: Accelerating resource allocation in wireless networks," \emph{IEEE Trans. Veh. Techn}., vol. 69, no. 1, pp. 958--970, Jan. 2020.
\bibitem{Hopfield}
G. A. Tagliarini \emph{et al}., ``Optimization using neural networks," \emph{IEEE Trans. Comput.}, vol. 40, no. 12, pp. 1347--1358, Dec. 1991.
\bibitem{ApproximateObjective}
G. Villarrubia \emph{et al}., ``Artificial neural networks used in optimization problems," \emph{Neurocomputing}, vol. 272, pp. 10--16, Jan. 2018.
\bibitem{GraphNN}
M. Eisen and A. Ribeiro, ``Optimal wireless resource allocation with random edge graph neural networks," \emph{IEEE Trans. Signal Process.}, vol. 68, pp. 2977--2991, Apr. 2020.
\bibitem{constraint_elimination}
M. R. Hestenes, ``Multiplier and gradient methods," \emph{J. Optim. Theory Appl}., vol. 4, pp. 303--320, Nov. 1969.
\bibitem{baseline}
H. Guo \emph{et al}., ``Weighted sum-rate maximization for reconfigurable intelligent surface aided wireless networks", {\em IEEE Trans. Wireless Commun.}, vol. 19, no. 5, pp. 3064--3076, May 2020.
\end{thebibliography}
\end{document}